\documentstyle[multicol,aps,prbbib,epsf]{revtex}

\newcommand{\lsim}
{\ \raise.35ex\hbox{$<$}\kern-0.75em\lower.5ex\hbox{$\sim$}\ }
\begin{document}
\draft
\title
{Theoretical study of quasiparticle states
near the surface of a quasi-one-dimensional organic superconductor
$(\mbox{TMTSF})_{2}\mbox{PF}_6$}
\author{Y. Tanuma \cite{Okayama}}
\address{Department of Applied Physics,
Nagoya University, Nagoya 464-8063, Japan}
\author{K. Kuroki}
\address{Department of Applied Physics and
Chemistry, The University of Electro-Communications, 1-5-1 Chofugaoka,
Chofu-shi, Tokyo 182-8585, Japan}
\author{Y. Tanaka}
\address{Department of Applied Physics,
Nagoya University, Nagoya 464-8063, Japan}
\author{S. Kashiwaya}
\address{Electrotechnical Laboratory, 
1-1-4 Umezono, Tsukuba, 305-0045, Japan}
\par
\date{\today}

\maketitle
\begin{abstract}
Quasiparticle states near the surface of 
a quasi-one-dimensional organic superconductor 
$(\mbox{TMTSF})_{2}\mbox{PF}_6$
are studied based on an extended Hubbard model
on a quasi-one dimensional lattice at quarter-filling.
Three types of pairing symmetries, (i) $p$-wave, (ii) $d$-wave,
or (iii) $f$-wave are assumed. 
The resulting surface density of states
has characteristic features for each pairing symmetry:
i) a zero-energy peak (ZEP) in a U-shaped structure,
ii) a V-shaped structure without ZEP,
and 
iii) a ZEP in a V-shaped structure.
From these results, we propose that the tunneling spectroscopy 
serves as a strong method to identify the 
the pairing symmetry in 
$(\mbox{TMTSF})_{2}\mbox{PF}_6$. 
\end{abstract}
\par
\widetext
\begin{multicols}{2}
In recent years,
pairing symmetry in various unconventional superconductors, 
such as the high-$T_{c}$ cuprates, heavy fermion systems,
$\mbox{Sr}_{2}\mbox{RuO}_{4}$, organic superconductors, and so on,
has been extensively studied 
both experimentally and theoretically.\cite{Scala,Sigrist,Kashiwaya}
In particular, 
quasi-one-dimensional (Q1D) 
organic superconductors $(\mbox{TMTSF})_{2}\mbox{X}$
(X=PF$_6$, ClO$_4$, etc.) have recently attracted much attention
as a possible {\em spin-triplet} superconductor.
Experimentally, 
the observation of a large critical magnetic field $H_{c2}$
exceeding Pauli paramagnetic limit\cite{Lee97}, as well as
an unchanged Knight shift across $T_c$,\cite{Lee01}
strongly suggest spin-triplet pairing.
As for the orbital part of the pair wave function,
the presence of nodes in the pair potential on the 
Fermi surface has been suggested from 
NMR measurements for $(\mbox{TMTSF})_{2}\mbox{ClO}_4$\cite{Takigawa}
and $(\mbox{TMTSF})_{2}\mbox{PF}_6$,\cite{Lee01}
which exhibit the absence of Hebel-Slichter peak as well as a 
power-law decay of $T_{1}^{-1}$ below $T_c$.
On the other hand, a thermal conductivity
measurement has suggested the absence of nodes on the Fermi surface
in $(\mbox{TMTSF})_{2}\mbox{ClO}_4$.\cite{Belin}
\par
Theoretically, several previous studies
have proposed a triplet $p$-wave pairing state,
\cite{Abrikosov,Hasegawa,Lebed}
for which the nodes of the pair potential do not intersect the 
Fermi surface.
On the other hand, a spin-singlet $d$-wave-like
pairing mediated by spin fluctuations has been proposed
by several authors.\cite{Shimahara,KA,Kino}
This is because superconductivity lies right next to the 
$2k_{\rm F}$ spin density wave (SDW) phase in the pressure-temperature phase
diagram.\cite{Greene}
Moreover, one of the present authors have recently proposed\cite{Kuroki} 
that triplet $f$-wave-like pairing may dominate over $d$- and $p$-wave 
in $(\mbox{TMTSF})_{2}\mbox{PF}_6$ due to a combination
of Q1D Fermi surface, coexistence of $2k_{\rm F}$ SDW
and $2k_{\rm F}$ charge density wave (CDW) suggested from diffuse X-ray
scattering,\cite{Pouget,Kagoshima} and an anisotropy in the spin fluctuations.
\par
Thus, the situation is not settled either experimentally or 
theoretically. The purpose of the present study is 
to propose an experimental method to determine which one of 
the pairing symmetries is realized in $(\mbox{TMTSF})_{2}\mbox{PF}_6$.
\par
Now, for the high-$T_{c}$ cuprates, 
which has a singlet $d$-wave pair potential,
it has been clarified that the internal phase causes
a drastic interference effect
in the quasiparticle states
near surfaces or interfaces, enabling us to 
detect the sign change in the pair potential.
Namely, a zero-energy bound state (ZES)
at a (110) surface of a $d$-wave superconductor
reflects the sign-change of the effective pair potential
in the process of the reflection
of quasiparticle at the surface \cite{Hu}.
The formation of ZES results in a peak in
the surface density of states (SDOS)
at the Fermi energy (zero-energy)
and manifests itself as a so-called zero-bias conductance peak (ZBCP)
observed in scanning tunneling spectroscopy,
\cite{Tanaka1,Geerk,Alff,Wei,Wang}
which is considered as a strong evidence for a sign change 
in the pair potential.
\par
Recently, Sengupta $et$ $al.$ have proposed that
the pairing symmetry in $(\mbox{TMTSF})_{2}\mbox{X}$ \cite{Sengupta}
can be determined from the presence/absence of the ZES on the surface.
Although their study points out an important aspect,
their argument is mainly restricted to the
absence/presence of ZEP, from which the $p$-wave and $f$-wave pairings 
cannot be distinguished. 
In fact, as we shall see, one has to look into the overall line
shape of the SDOS to distinguish $p$- and $f$-wave pairings.
Since the detailed line shape of the SDOS is significantly influenced 
by the actual shape of the Fermi surface, we have to consider 
a more realistic lattice structure, in which the 
quasi one-dimensionality (warping) of
the Fermi surface is taken into account.
\par
In order to meet this requirement, we consider an  
extended Hubbard model on a Q1D lattice at quarter-filling,
extending the previous study on a 2D square lattice.\cite{Tanuma1}
We concentrate on $(\mbox{TMTSF})_{2}\mbox{PF}_6$  because 
there is no complexity (unit cell doubling) 
due to anion ordering like in $(\mbox{TMTSF})_{2}\mbox{ClO}_4$.
\cite{Yoshino}
Three types of physically plausible pairings
(i) triplet '$p$-wave' (ii) singlet '$d$-wave' 
and (iii) triplet '$f$-wave' are studied.
The spatial dependence of the pair potentials
is determined self-consistently, and
the SDOS is calculated
using the self-consistently determined  pair potentials.
We propose from the calculation results that the 
quasiparticle tunneling spectroscopy should serve as a strong method 
to identify the pairing symmetry in $(\mbox{TMTSF})_{2}\mbox{PF}_6$.
\par
%
%
The extended Hubbard model is given as 
\begin{eqnarray}
\label{Hami01}
    {\cal H}=
 &-& \sum_{\langle {\bf i,j} \rangle_{a},\alpha}
  \left (t_{a}c^{\dagger}_{{\bf i},\alpha} c_{{\bf j},\alpha}
 + {\rm H.c.} \right )
\nonumber \\
 &-& \sum_{\langle {\bf i,j} \rangle_{b},\alpha}
  \left (t_{b}c^{\dagger}_{{\bf i},\alpha} c_{{\bf j},\alpha}
 + {\rm H.c.} \right )
\nonumber \\
  &-& \frac{V}{2}\sum_{({\bf i,j})_{m},\alpha,\beta}
      c^{\dagger}_{{\bf i},\alpha} c^{\dagger}_{{\bf j},\beta}
      c_{{\bf j},\beta} c_{{\bf i},\alpha}
 -\mu \sum_{{\bf i},\alpha}
      c_{{\bf i},\alpha}^{\dagger} c_{{\bf i},\alpha},
\end{eqnarray}
where
$c_{{\bf i}\alpha}$ [$c^{\dagger}_{{\bf i}\alpha}$]
is the annihilation [creation]
operator of an electron with spin $\alpha=\uparrow,\downarrow$
at site ${\bf i}=(i_{a},i_{b})$.
Here $t_{a}[_{b}]$ is the hopping integral,
and
$\langle {\bf i,j} \rangle_{a}[_{b}]$
stands for summation over nearest neighbor pairs
in the $a[b]$-axis direction, respectively.
$V$ is the inter-electron potential 
between sites 
separated by $m$ lattice spacings in the $a$-direction,
and
$({\bf i,j})_{m}$ represents summation
over pairs of sites separated by $m$ lattice spacings.
$m$ depends on the choice of the pairing considered.
We choose $t_{b}/t_{a}=0.1$  
in order to take into account the Q1D Fermi surface
of $(\mbox{TMTSF})_{2}\mbox{PF}_6$,
which is open in the $k_{b}$-direction.
The chemical potential $\mu$ 
is determined so that the band is quarter-filled. 
\par
\begin{figure}[htb]
\begin{center}
\epsfxsize=8cm
\epsfysize=6cm
\centerline{\epsfbox{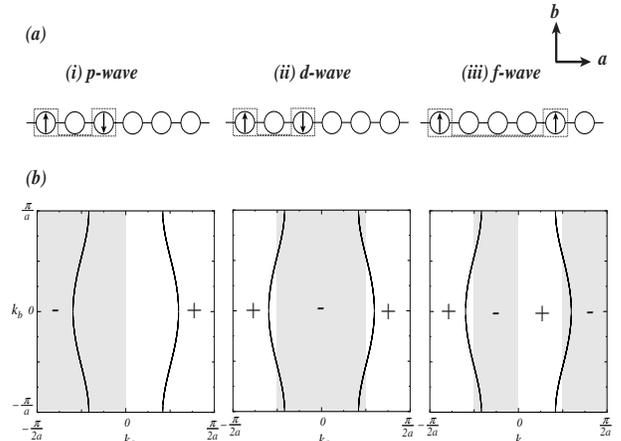}}
\caption{(a) An illustration of Cooper 
pairing in real space 
and (b) the shape of Fermi surface for $t_{b}/t_{a}=0.1$
at quarter-filling and the pair potential for
(i) triplet `$p$-wave' ($m_{p}=2$),
(ii) singlet `$d$-wave' ($m_{d}=2$),
and
(iii) triplet `$f$-wave' ($m_{f}=4$).
In (a), the pairs are depicted by dashed lines, and
in (b) $+$ $(-)$ denotes the region where 
the sign of the pair potential is
positive (negative).}
\label{fig1}
\end{center}
\end{figure}

By applying a mean-field approximation,
$\Delta^{\alpha \beta}_{{\bf ij}}
= \frac{V}{2} \langle c_{{\bf i}\alpha}c_{{\bf j}\beta} \rangle$
is introduced, which
represents
the superconducting pair potential for pairs formed
by $\alpha$-spin electron on the ${\bf i}$th site and 
$\beta$-spin electron on the ${\bf j}$th site.
We assume that $\Delta_{{\bf ij}}^{\alpha \beta}$
is proportional to $\delta_{i_b,j_b}$, where $i_b$ and $j_b$ are 
coordinates in the $b$-direction. Thus 
the unit cell contains $N_{\rm L}$ sites in the $a$-direction
and one site in the $b$-direction.
We consider three pairing symmetries 
shown in Fig.~\ref{fig1}. 
Namely,
(i) $p$-wave: $S_b=0$ ($S_b$ is the $b$-component of the total spin
of a pair) triplet pairing between sites separated by 2
lattice spacings ($m_{p}=2$). 
This is a `$p$-wave' pairing 
because the pair potential has a $2\Delta_{p} \sin 2k_{a}a$,
$k$-dependence in the bulk state, so that the pair potential 
changes its sign as $+-$ along the Fermi surface (see Fig.~\ref{fig1}).
(ii) $d$-wave: singlet pairing between sites separated by 2 
lattice spacings ($m_{d}=2$). 
This is a `$d$-wave' pairing 
in the sense that the pair potential changes 
its sign as $+-+-$ along the Fermi surface due to
its $2\Delta_{d} \cos 2k_{a}a$ $k$-dependence in the bulk state.
(iii) $f$-wave: $S_b=\pm 1$ triplet pairing between sites separated 
by 4 lattice spacings ($m_{f}=4$).
This is an `$f$-wave' pairing 
in the sense that the pair potential changes 
its sign as $+-+-+-$ along the Fermi surface 
due to its $2\Delta_{f} \sin 4k_{a}a$ 
$k$-dependence in the bulk state.
These three pairings are physically plausible
in the sense that they are consistent with the spin alignment of the 
$2k_{\rm F}$ SDW phase of
a quarter filled $(k_{\rm F}=\frac{\pi}{4a})$ system 
with an easy axis in the $b$-direction.\cite{Kuroki} 
In other words, the $2k_{\rm F}$ spin fluctuations 
can favor pairing in these channels.
\par
In order to represent $\Delta_{{\bf i}{\bf j}}^{\alpha \beta}$
in a more convenient way, 
we introduce a new coordinate $j$ along the $a$-direction.
The original coordinate ${\bf j}$ is
represented as ${\bf j}=(j,j_{b})$ with
$j=1,\cdots ,N_{\rm L}$.
In the $b$-direction,
we assume $N_{b}$ unit cells, and the electrons
are Fourier transformed as
$C_{j\alpha}(k_{b}) = \sum_{j_{b}=1}^{N_{b}}
      c_{{\bf j}\alpha}e^{-ik_{b}j_{b}a}$,
and
$C_{j\beta}(-k_{b}) = \sum_{j_{b}=1}^{N_{b}}
      c_{{\bf j}\beta}e^{ik_{b}j_{b}a}$,
where
$-\pi/a < k_{b} \leq \pi/a$,
$k_{b}=\frac{2\pi}{N_{b}}n$ 
with $n$ being an integer.
\par
%
After the Fourier transformation,
the mean-field Hamiltonian becomes
\begin{eqnarray}
\label{HMF}
   &{\cal H}_{\rm MF}&
      = \sum_{k_{b},i,j}
    \left [
      \begin{array}{cccc}
        C_{i\uparrow}^{\dagger}(k_{b}) & C_{i\downarrow}^{\dagger}(k_{b})
      & C_{i\uparrow}(-k_{b}) & C_{i\downarrow}(-k_{b})
      \end{array}
    \right ]
\nonumber \\
&\times &
    \left [
      \begin{array}{cccc}
    H_{ij} (k_{b}) & 0 &
    \Delta_{ij}^{\uparrow\uparrow} &
    \Delta_{ij}^{\uparrow\downarrow} \\
    0 & H_{ij} (k_{b}) &
    \Delta_{ij}^{\downarrow\uparrow} &
    \Delta_{ij}^{\downarrow\downarrow} \\
    \Delta_{ji}^{*\uparrow\uparrow} &
    \Delta_{ji}^{*\downarrow\uparrow} &
    -H_{ji} (-k_{b}) & 0 \\
    \Delta_{ji}^{*\uparrow\downarrow} &
    \Delta_{ji}^{*\downarrow\downarrow} &
    0 & -H_{ji} (-k_{b})
      \end{array}
    \right ]
\nonumber \\
& \times &
    \left [
      \begin{array}{c}
        C_{j\uparrow}(k_{b}) \\
        C_{j\downarrow}(k_{b}) \\
        C_{j\uparrow}^{\dagger}(-k_{b}) \\
        C_{j\downarrow}^{\dagger}(-k_{b}) 
      \end{array}
    \right ],
\\
  & H_{ij}(k_{b})& =
    - \sum_{\pm} \left [ t_{a}\delta_{i,j \pm 1}
                + 2t_{b} \cos (k_{b}a) \delta_{i,j}
                 - \mu \delta_{i,j} \right ].
\end{eqnarray}
Here, for simplicity,
the off diagonal part
in Eq.(\ref{HMF}) is assumed
in the following forms.
For the $p$-wave state,
\begin{eqnarray}
\Delta_{ij}^{\uparrow\downarrow}=
\Delta_{ij}^{\downarrow\uparrow}=
\sum_{\pm} \Delta_{ij}^{p}\delta_{i,j \pm 2},
\quad
\Delta_{ij}^{\uparrow\uparrow}=
\Delta_{ij}^{\downarrow\downarrow}=0,
\end{eqnarray}
for the $d$-wave state,
\begin{eqnarray}
    \Delta_{ij}^{\uparrow\downarrow}
&=& -\Delta_{ij}^{\downarrow\uparrow}
 =  \sum_{\pm} \Delta_{ij}^{d}\delta_{i,j \pm 2},
\quad
\Delta_{ij}^{\uparrow\uparrow}=
\Delta_{ij}^{\downarrow\downarrow}=0,
\end{eqnarray}
and for the $f$-wave state,
\begin{eqnarray}
    \Delta_{ij}^{\uparrow\uparrow}
&=& \Delta_{ij}^{\downarrow\downarrow}
 =  \sum_{\pm} \Delta_{ij}^{f}\delta_{i,j \pm 4},
\quad
\Delta_{ij}^{\uparrow\downarrow}=
\Delta_{ij}^{\downarrow\uparrow}=0.
\end{eqnarray}
We have taken the total number of sites as $N_{\rm L}=500$
and $N_{b}=50$.
The value of the pair potential and the chemical potential  
in the bulk with $V/t_{a}=4.0$ are 
(i) $\Delta_{p}/t_{a}=0.280$, $\mu/t_{a}=1.39$,
(ii) $\Delta_{d}/t_{a}=0.164$, $\mu/t_{a}=1.39$,
and
(iii) $\Delta_{f}/t_{a}=0.244$, $\mu/t_{a}=1.38$, respectively.
\par
%
In the actual numerical calculation,
the above Hamiltonian $\cal{H}_{\rm MF}$
is diagonalized by Bogoliubov transformation
\cite{Tachiki} given by
$C_{i\alpha}(k_{b}) =
   \sum_{\nu}{\cal U}_{i,\nu}\gamma_{\nu}(k_{b})$,
and
$C_{j\beta}(-k_{b})=
   \sum_{\nu}\gamma^{\dagger}_{\nu}(k_{b})
{\cal U}^{\ast}_{N_{{\rm L}}+j,\nu}$,
where $\nu$ is the index which specifies the eigenstates.
Then, the mean-field Hamiltonian described
in Eq.~(\ref{HMF}) is rewritten as
${\cal{H}}_{\rm MF}=\sum_{k_{b},\nu}E_{\nu}(k_{b})
\gamma_{\nu}^{\dagger}(k_{b})\gamma_{\nu}(k_{b})$,
where the operator $\gamma_{\nu}(k_{b})$ satisfies
the fermion's anticommutation relation.
The spatial dependence of the pair potential with
$l$-wave pairing symmetry is determined
self-consistently as
\begin{eqnarray}
\label{SCF}
  \Delta_{j,j \pm m_{l}}^{l}
    = \frac{V}{2} \sum_{k_{b},\nu}
    {\cal U}_{j \pm m_{l},\nu} {\cal U}^{\ast}_{N_{{\rm L}}+j,\nu}
    \left \{ 1-f[E_{\nu}(k_{b})] \right \},
\end{eqnarray}
where $f[E_{\nu}(k_{b})]$
denotes the Fermi distribution function.
The procedure is iterated 
until the pair potential $\Delta_{ij}^{l}$ 
is obtained fully self-consistently. 
%
We calculate the SDOS
using the pair potential determined self-consistently.
In order to compare our theory with
scanning tunneling microscopy (STM) experiments,
we assume that the STM tip is metallic with a flat density of states (DOS),
and that the tunneling probability is finite
only for the nearest site from the tip. 
This assumption has been verified 
through the study of tunneling conductance of unconventional 
superconductors. This is because the magnitude of the tunneling probability 
of an electron is sufficiently low in the actual STM experiments. 
The resulting  tunneling conductance spectrum converges to
the normalized SDOS \cite{Kashiwaya}
\begin{eqnarray}
   \bar{\rho}(E) &=&
   \frac{
   \displaystyle{
   \int ^{\infty}_{-\infty}{\rm d}\omega \rho_{1,{\rm S}}(\omega)
   {\rm sech}^{2}(\frac{\omega - E}{2k_{\rm B}T})}}
   {
   \displaystyle{
   \int ^{\infty}_{-\infty}{\rm d}\omega \rho_{{\rm N}}(\omega)
   {\rm sech}^{2}(\frac{\omega + 2\Delta_{l}}{2k_{\rm B}T})}} \\
   \rho_{1,{\rm S}}(\omega)
      & = &2 \sum_{k_{b}} \sum_{\nu} |{\cal U}_{1,\nu}|^{2}
      \delta\{\omega-E_{\nu}(k_{b})\}.
\end{eqnarray}
at low temperatures,
where $\rho_{i,{\rm S}}(\omega)$ denotes the SDOS
at the $i$-th site from the surface in the superconducting state
and $\rho_{{\rm N}}(\omega)$
denotes the DOS in the normal state.
In this paper, $\rho_{{\rm N}}(\omega)$ is obtained from
the DOS at the $N_{{\rm L}}/2$-th site far
away from the surface.
\par
%
\begin{figure}[htb]
\begin{center}
\epsfxsize=8cm
\epsfysize=11cm
\centerline{\epsfbox{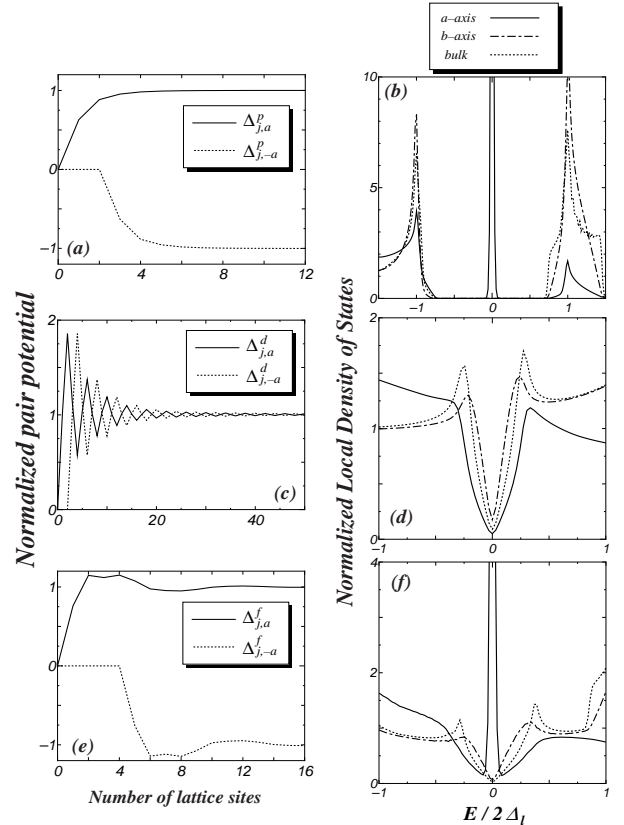}}
\caption{
The left panels are the spatial dependences 
of the pair potentials along the $a$-axis
near the surface in the $l$-wave state
($l=p,d,f$),
and the right panels show
the SDOS at the surface normal to the $a$- or $b$-axis 
along with the bulk density of states.}
\label{fig2}
\end{center}
\end{figure}
%
The obtained spatial dependences
of the $p$-, $d$-, and $f$-wave pair potentials 
and the corresponding SDOS are plotted in Fig.~\ref{fig2}.
Since the spatial dependence of the pair potential is complex,
we define the following quantities 
given by
\begin{eqnarray}
\Delta_{j,\pm a}^{p} &\equiv&
{\rm Re}[\Delta_{j,j \pm 2}^{p}]/\Delta_{p},
        \quad
{\rm Im}[\Delta_{j,j \pm 2}^{p}]=0,
        \nonumber \\
\Delta_{j,\pm a}^{d} &\equiv&
{\rm Re}[\Delta_{j,j \pm 2}^{d}]/\Delta_{d},
        \quad
{\rm Im}[\Delta_{j,j \pm 2}^{d}]=0,
        \\ \nonumber
\Delta_{j,\pm a}^{f} &\equiv&
{\rm Im}[\Delta_{j,j \pm 4}^{f}]/\Delta_{f},
        \quad
{\rm Re}[\Delta_{j,j \pm 4}^{f}]=0.
\end{eqnarray}
to visualize the spatial dependences clearly.
The left panels of Fig.~\ref{fig2} is the obtained result
for the spatial dependence of the pair potential
near a surface normal to the $a$-axis,
and the right panels
show the SDOS at the surface normal to the $a$- or the $b$-axis
along with the bulk DOS.
\par
First, let us look into the results for the triplet $p$-wave
pairing state shown in the upper panels of Fig.~\ref{fig2}
[see Fig.~\ref{fig2}(a) and~\ref{fig2}(b)].
Since triplet Cooper pair is
formed between two electrons
with 2 lattice spacings, 
$\Delta^{p}_{j,a}=-\Delta^{p}_{j+2,-a}$ is satisfied. 
Both the magnitude of 
$\Delta^{p}_{j,a}$ 
and $\Delta^{p}_{j,-a}$
is suppressed near the surface
and approaches $1$ and $-1$
in the middle of superconductor, respectively.
As shown in Fig.~\ref{fig2}(b), 
the corresponding DOS
has a U-shaped gap structure 
similar to that of the conventional $s$-wave pairing 
due to the fact that the nodes of the pair potential 
do not intersect the Fermi surface.
The ZEP shows up in SDOS at the surface normal to $a$-axis 
due to the formation of ZES,
since an injected and reflected quasiparticle
feel different sign of the pair potential
\cite{Buchholtz}.
On the other hand, at the surface normal to the $b$-axis,
since an injected and reflected quasiparticle
feel the same pair potential,
the ZES is not formed and the resulting SDOS 
has no ZEP, resulting in a overall line shape 
similar to that of the bulk DOS.
These results for the $p$-wave pairing are consistent 
with those in Ref.~\onlinecite{Sengupta}.
\par
%
Next, we look into the corresponding quantities 
in singlet $d$-wave pairing case
[see Fig.~\ref{fig2}(c) and~\ref{fig2}(d)]. 
Since the pair is formed between sites separated 
by 2 lattice spacings,
$\Delta_{j,a}^{d}=\Delta_{j+2,-a}^{d}$ is satisfied.
The obtained spatial dependence of the pair potential
exhibits a atomic-scale spatial oscillation
near the surface and converges  to 
the bulk value toward the middle of the lattice.
These features are similar to the previous results for 
the extended Hubbard model on 
a 2D square lattice \cite{Tanuma1}.
The corresponding SDOS (and bulk DOS)
has a V-shaped 
structure due to the existence of nodes of the pair potential 
on the Fermi surface.
However,
since an injected and reflected quasiparticle 
feel the same pair potential both at the surfaces normal to
$a$- and $b$-axis, no ZEP appears at the SDOS
[see Fig.~\ref{fig2}(d)].
Due to the absence of the pair potential
along the $b$-axis,
the ZEP never appears for arbitrary orientation
of the surface which is strikingly different from the case of 
the high-$T_{c}$ cuprates \cite{Tanaka1}.
\par
%
Finally,
we move on to the case of the triplet $f$-wave pairing.
The $f$-wave pair 
[see Fig.~\ref{fig1}(a) (iii)]
is formed between sites separated by 
4 lattice spacings, so the resulting pair potential
satisfies $\Delta_{j,a}^{f}=-\Delta_{j+4,-a}^{f}$.
As seen from Fig.~\ref{fig2}(e),
the obtained pair potential has a complex spatial dependence
as compared to that of the $p$-wave pairing.
Comparing Fig.~\ref{fig2}(b) and Fig.~\ref{fig2}(f),
it can be seen that 
the $f$-wave pairing belongs to the same class as that of the 
$p$-wave pairing
{\em as far as the absence/presence of the ZEP is concerned,}
as has been pointed out in Ref.~\onlinecite{Sengupta}.
However, since the $f$-wave pair potential 
has nodes on the Fermi surface, the resulting SDOS (and bulk DOS)
has a V-shaped structure similar to that for the $d$-wave case
in sharp contrast with the case of $p$-wave pairing.
\par
In total, as summarized in Table~\ref{table1}, 
the $p$-,$d$-,and $f$-wave pairings can be clearly distinguished from
the combination of the overall line shape of the SDOS and
the presence/absence of the $a$-axis ZEP. 
\par
In summary,
we have studied the quasiparticle SDOS of
an organic superconductor $(\mbox{TMTSF})_{2}\mbox{PF}_6$
based on an extended Hubbard model on a Q1D lattice at quarter filling.
The non-local feature of the pair potential
and the atomic-scale geometry of the surface 
are explicitly taken into account in the present calculation.
Three types of pairing symmetries,
(i) $p$-wave, (ii) $d$-wave,
and (iii) $f$-wave have been considered.
The calculation results suggest that we can clearly distinguish the 
present three pairing symmetries from tunneling spectroscopy.
We believe our theoretical prediction 
can be verified experimentally in the near future. 
It is an interesting future problem to investigate how our results 
will be modified for $(\mbox{TMTSF})_{2}\mbox{ClO}_4$,
in which a unit cell doubling due to anion ordering takes place.
\cite{Yoshino}
It is also an interesting future problem to study Josephson 
effect in Q1D organic superconductors 
since it has been clarified in the previous studies 
that Josephson effect are crucially influenced 
by unconventional pair potentials.\cite{Tanakaj96}  
\par
First author (Y. T.) acknowledges the financial support
of Research Fellowships of Japan Society for
the Promotion of Science (JPSJ) for Young Scientists.
K.K. acknowledges Hideo Aoki for discussions and 
pointing out Ref.\onlinecite{Yoshino}.
The computational aspect of this work has been performed at the
facilities of the Supercomputer Center,
Institute for Solid State Physics,
University of Tokyo and the Computer Center.
%
%

\begin{table}
\caption{Surface density of states (SDOS)
for $p$,$d$, and $f$-wave pairings.
\label{table1}}
\begin{tabular}{cc}
Symmetry & SDOS  \\ \hline
$p$-wave & U-shaped + $a$-axis ZEP  \\
$d$-wave & V-shaped + No ZEP  \\
$f$-wave & V-shaped + $a$-axis ZEP 
\end{tabular}
\end{table}
\end{multicols}
\end{document}